\documentclass[11pt, oneside]{article}   	
\usepackage{geometry}                		
\usepackage{graphicx}				
\usepackage{amssymb}

\begin{document}
\title{A possible experimental test of the nonlinear phononics interpretation of light-induced superconductivity}
\author{M. Altarelli \\  Max Planck Institute for Structure and Dynamics of Matter \\ 22761 Hamburg, Germany}
\date{22 February 2019}							
\maketitle 
\begin{abstract}
Experimental evidence for a transient enhancement of the superconducting critical temperature in presence of an intense THz or IR pump pulse was ascribed to nonlinear phononics effects. Here I introduce a simple phenomenological Ginzburg-Landau model of this phenomenon, to explore further consequences and possible experimental tests of this interpretation. Upon cooling below $T_{c}$ in the absence of THz pumping, both an abrupt softening of a Raman-active mode frequency and a spontaneous lattice distortion, growing linearly with $(T_{c}-T)$, are predicted to occur. Numerical estimates for $YBa_{2}Cu_{3}O_{6+x}$ indicate that the frequency softening should be easily observable, whereas the lattice distortion may be too small. Comparison with Raman experiments for $YBa_{2}Cu_{3}O_{6+x}$ is far from conclusive; however very large (up to 18 \%) phonon frequency softening below $T_{c}$, with behaviour strikingly similar to the predictions of the present model, was observed over 20 years ago in $HgBa_{2}Ca_{3}Cu_{4}O_{10+x}$; its explanation has been controversial. Light induced superconductivity was never investigated in this material and it may be of interest to explore if it is present and connected, via the mechanism discussed here, to the observed anomalous phonon behaviour. 
\end{abstract}

\section{Introduction}
 
 The purpose of this note is to explore further possible experimental tests for the non-linear lattice dynamics explanation \cite{MF, Subedi, Mankowsky} of light-enhanced superconductivity in \linebreak $YBa_{2}Cu_{3}O_{6+x}$  \cite{Fausti, Hu2014, Kaiser2014}. According to this explanation, one (or more) optical phonon  $A_{1g}$-symmetric mode coordinate, $Q$, is pushed off its equilibrium value by excitation of an infrared-active mode $Q_{IR}$, due to a nonlinear coupling that, according to Mankowsky's \textit{et al.} arguments is of the form $ -a_{1,2} Q Q_{IR}^{2}$. Indeed, when this term is added to the harmonic potentials and the kinetic energies of the $ Q_{IR}, Q$ oscillators, the Barbanis Hamiltonian \cite{Barba} is obtained, which has been first investigated in the context of celestial mechanics, then in studies of quantum ergodicity and molecular physics \cite{Nordholm},\cite{MaluendBr}, \cite{Maluendes}; and it turns out that an accurate approximation to the exact eigenfunctions of this Hamiltonian is given by eigenfunctions of the form  \cite{MaluendBr}, \cite{Maluendes}:

\begin{equation}
\Phi_{n_1,n_2} = N\phi_{n_1}(aQ+b) \phi_{n_2}(a_{IR}Q_{IR})
\end{equation}

in terms of harmonic oscillator eigenfunctions for scaled oscillator coordinates, and, for the coordinate $Q$, with a displaced oscillator coordinate; the scaling and displacement terms $a, a_{IR}$ and $b$ can be determined variationally or within a SCF approach for a given value of $a_{1,2}$. $N$ is a normalization constant.
When the $Q$ lattice coordinate is pushed to an off-equilibrium value by excitation of the infrared mode by a pump pulse, evidence that could indicate a transient state with a substantial increase of $T_{c}$ \cite{Fausti, Hu2014, Kaiser2014} has been reported. 

If a given phononic coordinate affects the electronic states at the Fermi energy and considerably modifies $T_{c}$, as argued in Ref \cite{Mankowsky}, one could suspect that changes of the electronic states at the Fermi energy, such as the usual onset of equilibrium superconductivity when cooling below $T_{c}$, with opening of a gap, should have a corresponding effect on the phononic coordinate. This paper investigates the possibility that this could lead to observable effects that validate the nonlinear phononics interpretation of superconductivity.
  
To this aim, we adopt a very simple-minded Ginzburg-Landau description of the superconducting transition \cite{Lifshitz}, including an additional dependence of the critical temperature $T_c$ on the $Q$ lattice coordinate. Reasoning based on this phenomenological description implies that, even in the absence of the pump, when the system is cooled below $T_c$, a softening of this phonon frequency and a spontaneous distortion of the lattice should occur. An attempt to estimate the sizes of the softening  and of the distortion is carried out, at least roughly, to determine whether they could be observable;  measured experimental properties of $YBa_{2}Cu_{3}O_{6+x}$ are used to infer the values of the parameters entering the Ginzburg-Landau free-energy. It turns out that, for plausible values of the parameters, the frequency shift should be observable, whereas the lattice distortion could be harder to detect.
 
 \section{Ginzburg-Landau functional for the free-energy}
 In the phenomenological Ginzburg-Landau theory, the free-energy per unit volume of a superconductor in the vicinity of $T_c$  is written as a functional of the order parameter $\Psi$; in the simplest case of a uniform superconductor without external fields,  it takes the form:
 \begin{equation}
 F (T, \mid \Psi \mid) = a \times (T-T_c){\mid \Psi \mid}^{2} + b {\mid \Psi \mid}^{4}
 \end{equation}
 where the coefficients $ a, b$ are positive. Looking for a minimum of this functional as a function of $ \mid \Psi \mid$ one immediately finds $\mid \Psi \mid =0$ for $T>T_c$, and, for  $T<T_c$ the value
 \begin{equation}
 {\mid \Psi \mid}^{2}= \frac{a (T_c-T)}{2b}
 \end{equation}
 
 The free energy value corresponding to this minimum (the condensation energy) is evaluated plugging Eq. (3) into (2):
 
 \begin{equation}
 F_{min}(T)= a(T-T_{c}) \frac{a(T_{c} - T)}{2b} + b\frac{a^{2} (T_{c} - T)^{2}}{4b^2} = \frac{-a^{2} (T_{c} - T)^{2}}{4b}
 \end{equation}
 
 It can be argued that this mean-field type of formulation is oversimplified for the cuprates, with their short coherence length, and their symmetry lower than cubic. But for the sake of simplicity we shall stick to this description. The observation of very similar phenomena in totally different compounds, such as $K_{3}C_{60}$ \cite{Mitrano}, supports a rather generic approach. \newline 
 We further assume that an optical phonon coordinate $Q$, with a vanishing expectation value in the equilibrium state above $T_c$, is affecting the transition temperature, when it takes a non vanishing amplitude. A simple description of this state of affairs can be given by modifying the free-energy functional in the following way:
 
 \begin{equation} 
  F (T, \mid \Psi \mid, Q) = (1/2) \rho k Q^{2} + a [T-T_{c}- \Theta_{c} (Q)]{\mid \Psi \mid}^{2} + b {\mid \Psi \mid}^{4}
\end{equation}

Here, the first term on the right-hand side is the elastic energy (per unit volume) corresponding to the lattice distortion, with $\rho$ being the number density of deformed bonds and $k = m^{\ast} \omega^{2}$ the product of effective mass and square frequency of the $Q$ mode. The second term expresses an increase of the  critical temperature  to $T_{c}+ \Theta_{c} (Q)$, when $Q$ is nonvanishing. The function $\Theta_{c} (Q)$ reflects the complex physics that leads to a $T_c$ enhancement in a distorted lattice. According to the arguments in ref. \cite{MF, Subedi}, the lattice distortion induces a modification of the orbital character and  density of electronic states at the Fermi energy, leading to the $T_c$ enhancement. Lacking a microscopic theory, the functional relationship between $Q$ and $T_c$ remains out of our reach. \newline
Once again, a word of caution on the questionable aspects of this approach is necessary. As it was stated above, the Ginzburg-Landau theory is limited to $T$ near $T_c$. Just shifting the critical temperature keeping the same functional form and $a,b$ parameters is really justified if the shift satisfies
\begin{equation}
\Theta_{c} (Q) << T_c
\end{equation}

This restriction certainly does not apply to the situation described in ref. \cite{Kaiser2014}, which, taken at face value, implies e.g. a shift of some $250 K$ from $50 K$ to room temperature for the critical temperature of $YBa_{2}Cu_{3}O_{6.5}$ after a THz pulse. So one must understand Eq. (5) as describing the "small Q" regime, where Eq. (6) is satisfied, and not the regime of intense THz pumping that the experiments access. The quotation marks around "small Q" are justified by the circumstance that, according to ref. \cite{Mankowsky}, the $250 K$ rise in $T_c$ corresponds to a $0.63 \%$ elongation of the intra-Cu bilayer distance, which in other contexts one could call a small distortion.\newline
 
In order to make progress we make a linearization assumption for $
\Theta_{c}(Q)$ i.e. assume that in some range of Q 

\begin{equation} 
\Theta_{c}(Q) \simeq cQ
\end{equation} 
(remember that $\Theta_{c}(Q=0) =0$), where c is a coefficient to be determined. This is very likely the case for a sufficiently small Q range (the exception being if, for some reason, $(\partial\Theta_{c}/ \partial Q)_{Q=0}$ vanishes). Within this assumption Eq. (3) becomes:
\begin{equation}
F (T, \mid \Psi \mid, Q) = (1/2) \rho k Q^{2} + a [T-T_{c}- cQ]{\mid \Psi \mid}^{2} + b {\mid \Psi \mid}^{4}
\end{equation}

Additional support for this form of the free-energy functional comes from the discussion of the coupling of the strain tensor to the superconducting order parameter by Joynt and Rice \cite{Joynt}, Ozaki \cite{Ozaki} and Millis and Rabe \cite{MillisRabe} who, starting from a simple description of the condensation energy in terms of the band structure and the gap function, suggest coupling terms linear in the strain and quadratic in the order parameter (for a summary and extensive information on a generalization of the argument for anisotropic superconductors, see Ref. \cite{Sigrist}). This is exactly the character of the additional term in Eq. (8), $-cQ|\Psi|^{2}$. Although the strain tensor is representative of acoustical deformations, and here $Q$ represents an optical deformation, this should not modify the argument.

If a certain value of $Q\neq 0$ is imposed by the external THz pump, we can minimize the free-energy as in the previous case and get:
\begin{equation} 
 {\mid \Psi \mid}^{2}= \frac{a(T_c + cQ -T)}{2b}
\end{equation}

and
 \begin{equation} 
F_{min}(T, Q) = (1/2) \rho k Q^{2} - \frac{a^{2} (T_{c}+ cQ - T)^{2}}{4b} 
  \end{equation}
  
 Comparing to the $Q=0$ case, one finds:
  \begin{eqnarray} 
F_{min}(T, Q)  - F_{min}(T, Q=0) & = & (1/2) \rho k Q^{2} - \frac{a^{2} (T_{c}+ cQ - T)^{2}}{4b} +\frac{a^{2} (T_{c} - T)^{2}}{4b} \nonumber \\
 & = & (1/2) (\rho k - \frac{a^{2} c^{2}}{2b}) Q^{2} - \frac{a^{2} c (T_{c} - T)}{2b} Q 
  \end{eqnarray}
 
 or equivalently:

 \begin{equation} 
F_{min}(T, Q)   =  F_{min}(T, Q=0) + (1/2) (\rho k - \frac{a^{2} c^{2}}{2b}) Q^{2} - \frac{a^{2} c (T_{c} - T)}{2b} Q 
  \end{equation}

\section{Consequences in the absence of external pump pulse}

In this Section we shall consider the situation in which no external pump is present, to examine whether the $F_{min}(T, Q))$, Eq. (12) may lie below  $ F_{min}(T, Q=0)$ so that a spontaneous distortion takes place. Before proceeding in an analytical fashion to this simple task, inspection of Eq. (12) immediately reveals that the search for the minimum of  $F_{min}(T, Q))$ makes sense only when the following inequality holds:
\begin{equation}
\rho k  > \frac{a^{2} c^{2}}{2b}
\end{equation}

because in the opposite case both the quadratic and the linear term in the r.h.s. of Eq. (11) are negative and grow in size without limit with growing $Q$. Violation of the inequality (13) would signal a \textit{lattice instability} triggered by the very strong increase of the critical temperature with the lattice distortion. Manifestly this \textit{does not} happen in YBCO, and this is a first constraint on numerical values of the parameters $a,b,c,\rho$ and $k$, that we shall discuss in the next Section. As a matter of fact the lattice stability implies that $c^{2}$ is bound to stay below the value $c_{max}^{2}$:
\begin{equation}
c^{2} < c_{max}^{2} \equiv \frac{2b\rho k}{a^{2}}
\end{equation}

In terms of $c_{max}^{2}$, it is easy to express the softening of the bare frequency $\omega$ as a function of $c$ in the form:
\begin{equation}
\omega_{soft}^{2} = \omega^{2} (1-\frac{c^{2}}{c_{max}^{2}})
\end{equation}
In the case that $c$ is below this limit, corresponding to a softening of the phonon mode (reduction of the spring constant),  it makes sense to  minimize  Eq. (12) as a function of $Q$, a simple exercise that yields the equation:
\begin{equation}
(\rho k - \frac{a^{2} c^{2}}{2b}) Q_{min}  - \frac{a^{2} c (T_{c} - T)}{2b} = 0,
\end{equation}

and therefore 

\begin{equation}
Q_{min}(T) = \frac {a^{2}c (T_{c} -T)} {2b\rho k-a^{2} c^{2}}.
\end{equation}

This can also be cast in an appealing form in terms of $c_{max}$:
\begin{equation}
\frac{Q_{min}}{T_{c}-T} = \frac{c}{c_{max}^{2} - c^{2}}
\end{equation}

The corresponding free-energy value is, after some simple algebra:

 \begin{eqnarray}
 F_{min} (T, Q_{min}) & = & F_{min}(T,0) -\frac{a^{4}c^{2}(T_{c}-T)^{2}}{4b(2b\rho k -a^{2} c^{2})} \nonumber \\
& = & -\frac{a^{2} (T-T_{c})^{2}}{4b}[ 1 + \frac{c^{2}}{c_{max}^{2}- c^{2}}]
\end{eqnarray}

It is therefore easy to see that, provided Eq. (13) holds, two consequences of the $Q$ dependence of the effective critical temperature arise:

1. A softening of the phonon frequency (reduction of the "spring constant" in Eq. (11)), taking place below  $T_{c}$ in agreement with Eq. (15)

2. A spontaneous distortion from $Q=0$ to $Q_{min}(T)$ given by Eq. (18), linearly increasing in size as the temperature is lowered. 

Questions to address are whether the simple-minded treatment adopted here can lead to credible predictions, and whether the size of the predicted effects is possibly leading to observable effects. In order to explore the latter, one needs to establish numerical values of the model parameters in a realistic case.

\section{Numerical estimates for YBCO}

According to Ref. \cite{Mankowsky} the main effect of the THz pumping pulse is to generate a non-equilibrium elongation of the intra-bilayer distance, via non-linear phonon couplings. This elongation corresponds (ref. \cite{Mankowsky}, Extended Data Figure 3) to the modes labelled $A_{g}15$ and $A_{g}21$, with wave numbers $94.52 cm^{-1}$ and $125.19 cm^{-1}$. This assignment is not unambiguous, as more recent work by Fechner and Spaldin \cite{Spaldin}, also based on DFT calculations, suggests that the relevant $A_{g}$ mode is a different one with a substantially higher frequency, of order $470 cm^{-1}$. Nonetheless, just to get an order of magnitude estimate, let us take Mankowsky \textit{et al.}'s estimate and further simplify this to a single mode; to evaluate the parameter $k = m^{\ast} \omega^{2}$, let us assume a wavenumber of $\simeq 100 cm^{-1}$, corresponding to $\omega \simeq 2 \times 10^{13} s^{-1}$. The effective mass corresponding to the relative coordinate of two $Cu$ atoms is $(1/2) m_{Cu}\simeq (1/2) 63.5 amu$, where $1 amu = 1.66\times 10^{-24} g$ and $63.5$ is the isotope-averaged mass of a $Cu$ atom. One thus comes to 
\begin{equation}
k\simeq 2.1 \times 10^{4} g/s^{2}. 
\end{equation}

(Note that this purely indicative estimate of $m^{\ast} \omega^{2}$ is unchanged for a different, O-related phonon mode, if the effective mass is reduced by a factor 4 (from $Cu$ to $O$) and the frequency is increased by a factor of 2).

The density of Cu pairs facing each other across a bilayer in the ortho II structure of YBCO is one pair in a volume corresponding to $3.827\times 3.893\times 11.699 \AA ^{3}$ \cite{Calestani}, that establishes another parameter:
\begin{equation}
\rho \simeq 5.73 \times 10^{21}cm^{-3}.
\end{equation}

In order to obtain the value of the parameter $a^{2} / 2b$, which determines (together with the linearization coefficient $c$) the phonon softening, Eq. (15), and the lattice distortion, Eq. (18), we use a relation involving the specific heat jump per unit volume at $T_{c}$ (see Eq. (45.8) in ref.\cite{Lifshitz}):
\begin{equation}
(1/V)(C_{s} - C_{n}) = \frac{a^{2} T_{c}}{2b}
\end{equation}

Experimental determinations of the specific heat jump at $T_{c}$ are found in the literature. For $Y Ba_{2}Cu_{3} O_{6.5}$ Loram \textit{et al.} \cite{Loram} obtain for $\Delta\gamma= \gamma_{s}-\gamma_{n}$, where $\gamma= C/T$:

\begin{equation}
\Delta\gamma(T_{c}) \simeq 0.5 \frac {mJ}{gram-atom K^{2}}
\end{equation}

whereas W\"{u}hl \textit{et al.} \cite{Wuhl} measure 
\begin{equation}
\Delta\gamma(T_{c}) \simeq 10 \frac {mJ}{mole K^{2}}.
\end{equation}

To compare these two results it is useful to note that $ 1 gram-atom \equiv 1/(12+x) moles$ in $Y Ba_{2}Cu_{3} O_{6+x}$ \cite{Loram}, so that the value in Eq. (23) actually corresponds to:

\begin{equation}
\Delta\gamma(T_{c}) \simeq 6.2 \frac {mJ}{mole K^{2}}
\end{equation}

So an average gives an estimate of $8 mJ/(mole K^{2})$. The critical temperature of  $Y Ba_{2}Cu_{3} O_{6.5}$ is about $50 K$, and from Wikipedia \cite{Wiki} we learn that the density of the material is $6.3 g/cm^{3}$ and the molar volume is $666.2 g/mole$, so that we can find the required expression:

\begin{equation}
\frac {C_{s} - C_{n}}{V T_{c}} = \frac{a^{2}}{2b}= \frac{8.5\times 10^{-5}J}{cm^{3} K^{2}}
\end{equation}

Now all the ingredients are there to determine the numerical value of $c_{max}$. One obtains:

\begin{equation}
c_{max}^{2} =  \frac{2b\rho k}{a^{2}} =\simeq 1.4 \times 10^{23} K^{2}/cm^{2}
\end{equation}

or $c_{max} \simeq 3.8\times 10^{11} K/cm$. With the help of Eq. (15) and (18), we can now plot the lattice distortion and the softening of the frequency as a function of the linearization parameter $c$ (Figures 1  and 2).

\section{Preliminary comparisons with experiment}

The question whether these are observable effects is difficult to answer in the absence of a reasonable way to estimate the parameter $c$. If we take, on purely dimensional grounds, the ratio of the inferred increase of the critical temperature ($250 K$) and the measured elongation of the intralayer distance under strong $THz$ excitation, $2 pm$ \cite{Mankowsky}, we obtain a value ( $\simeq 120 K/pm$) exceeding $c_{max}$ by a factor of  3. If on a mere guess, we take a value of one half of $c_{max}$, this would result in a reduction of the phonon frequency by about $10 \%$, which should be detectable in Raman or in other experiments, especially because of its sudden onset as the temperature is lowered below $T_{c}$. The lattice distortion, which grows linearly with $T_{c} - T$, on the other hand, would be limited to $0.2 pm$  at $10 K$ below $T_{c}$, which is maybe below the detection limit (I do not feel comfortable about lowering the temperature further, as this would drive us outside of the region where Ginzburg-Landau arguments can possibly apply).

Coming to the Raman experiments on YBaCuO in the literature, a thorough study is reported by Limonov \textit{et al.} in Ref. \cite{Limonov2}. Their attention is focussed on a mode at $\simeq 340 cm^{-1}$, that among all Raman active $A_{1g}$ modes (in the orthorombic $D_{2h}$ classification) stands out for its larger distinct temperature dependence below $T_{c}$. In fact, a softening of over 2\% is shown for overdoped ($T_{c}=86 K$) and optimally doped ($T_{c}=93 K$)) samples, but much smaller for underdoped ($T_{c}=80 K$) ones. Their analysis is based on the theory of the phonon self-energies in a strong coupling superconductor as put forward by Zeyher and Zwicknagl \cite{ZZ}, modified for the anisotropic d-wave case by Nicol \textit{et al.} \cite{Nicol}. All other modes have much smaller frequency shifts (positive or negative) at and below $T_{c}$. According to the letter of the analysis of Zeyher and Zwicknagl, the mode is much more affected because in the tetragonal $D_{2h}$ point group classification this mode would be an "allowed" $B_{1g}$ mode, while the other considered modes are "forbidden" $A_{1g}$ (forbidden in the sense that selection rules would impose vanishing values of the matrix elements for the electron-phonon interactions affecting the phonon self-energy). This takes the point of view that the orthorombic distortions are very small, and disorder in the samples essentially washes their effects out, so that tetragonal concepts still apply. A further point to stress is that while the arguments of the present paper predict a sudden drop of the frequency at $T_{c}$, the observed temperature behaviour is rather steep but not really sudden. Finally, the mode in question is also not directly identifiable with the one singled out in Ref \cite{Spaldin}.

So the situation for YBaCuO is not encouraging, as far as comparison of the present paper with experiments is concerned. The mode identified by Mankowsky \textit{et al.} in Ref. \cite{Mankowsky} is not sensitive to temperature, and the one displaying much larger sensitivity may be interpretable by a different mechanism, i.e. in terms of self-energy effects. This could, among other possibilities, correspond to the fact that the parameter $c$ describing the electron lattice coupling in the small $Q$ regime happens to be very small.

Interestingly, experiments on cuprates with more than two layers, in particular $HgBa_{2}Ca_{3}Cu_{4}O_{10+x}$ \cite{Cardona}, a compound in which light-enhanced superconductivity was, as far as we know, never looked for,  are more in agreement with the predictions of the present analysis. For two modes of $A_{1g}$ symmetry in this tetragonal structure, with frequencies near $250$ and $390 cm^{-1}$ respectively, very large softenings (respectively 6\% and 18\%) are observed to occur rather abruptly in a $10-15 K$ interval below $T_{c} \simeq 120 K$. The interpretation of these experiments is still controversial. According to Munzar and Cardona \cite{Munzar}, the large phonon anomalies are related to the appearance, below $T_{c}$, in three- and four layers cuprates, of a Raman-active transverse Josephson plasma resonance; the charge fluctuations in the cuprate layers induced by the plasmon interact strongly with the charge distortions induced by the phonons, leading to a strong level repulsion and transfer of oscillator strength, with a corresponding enhancement of the Raman intensity for the softened phonon. Indeed, similar phonon anomalies are observed not only in the four-layer $HgBa_{2}Ca_{3}Cu_{4}O_{10+x}$ but also in the three-layer $Bi_{2}Si_{2}Ca_{2}Cu_{3}O_{10+x}$ system \cite{Limonov} with an $8\%$ softening of a Raman active phonon at $390cm^{-1}$. However Limonov \textit{et al.} observed that the intensity enhancement disappears when the incoming photon energy is increased from $\simeq 1.9 eV$ to a $2.2$ to $2.5 eV$ range, suggesting a resonance effect as the explanation. Following a theoretical suggestion by Venturini \textit{et al.} \cite{Venturini} that a Raman active $A_{1g}$ mode arises from collective spin excitations, Gallais \textit{et al.} \cite{Gallais} compare Raman spectra for the single layer $HgBa_{2}CuO_{4+x}$ and the three-layer $HgBa_{2}Ca_{2}Cu_{3}O_{8+x}$, and conclude that resonance behaviour and scaling with $T_c$ of the $A_{1g}$ modes are rather independent of the number of layers, and therefore do not support the Josephson plasmon explanation.

As far as $Bi_{2}Si_{2}Ca_{2}Cu_{3}O_{10+x}$ is concerned, however, Hu \textit{et al.} \cite{Hu-Nicoletti} report the lack of any signatures of light-enhanced critical temperature in this system. Therefore, whatever the reason for the phonon anomaly reported by Limonov \textit{et al.}, it is not related to the mechanism discussed here. 

\section{Conclusions} 

In this paper an attempt was made to explore possible experimentally verifiable consequences of the non-linear phononics interpretation of the phenomenon of light induced superconductivity. On the basis of a simple Ginzburg-Landau free-energy description, an abrupt softening below $T_c$ of  one (or more) Raman active phonon(s) was predicted. Inserting reasonable values for YBCO, it was concluded that the effect should be observable in this material. Comparison with existing experiments cannot be claimed to support this prediction, at least for the phonons that are identified as directly related to the $T_c$ enhancement. 

Observed phonon softenings in Bi (2223) and Hg (1234) cuprates with $3$ or $4$ $Cu-O$ layers are on the other hand quite similar to those predicted here. However, a recent search for light induced superconductivity in Bi (2223) gave a negative result \cite{Hu-Nicoletti}, while no search was carried out in Hg(1234).

On the basis of the presently existing data, the nonlinear phononics explanation cannot be directly and unambiguously supported by the arguments presented here. It may be of interest to explore the observability of a light induced enhancement of $T_c$ in Hg (1234) and in general to extend the search for the specific phonon modes driving this spectacular phenomenon.

 \section{Acknowledgments}
 
 I am greatly indebted to Andrea Cavalleri for his continuing interest in this work and for many valuable discussions. I am also grateful to Michael Fechner, Michael F\"{o}rst,  Andrew J. Millis, Angel Rubio and Michael Sentef for useful discussions.
 
\begin{figure}[b]
\includegraphics[width=15cm]{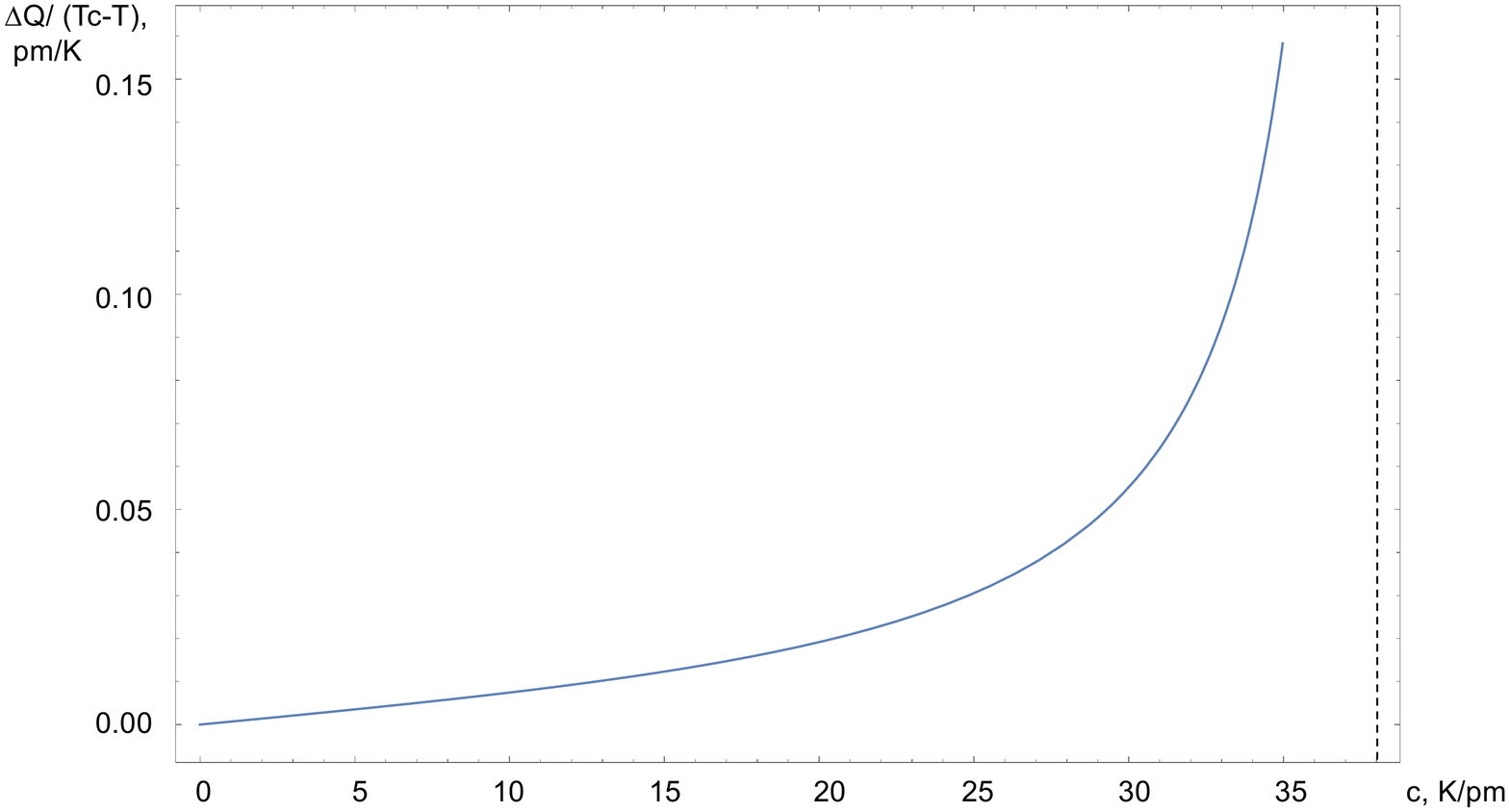}
\caption{\it{Lattice distortion in pm divided by the temperature below $T_{c}$ in K, as a function of the linearization parameter c, in $K/pm$. The vertical dashed line marks the $c_{max}$ value, $38 K/pm$}}
\end{figure}

\begin{figure}[b]
\includegraphics[width=15cm]{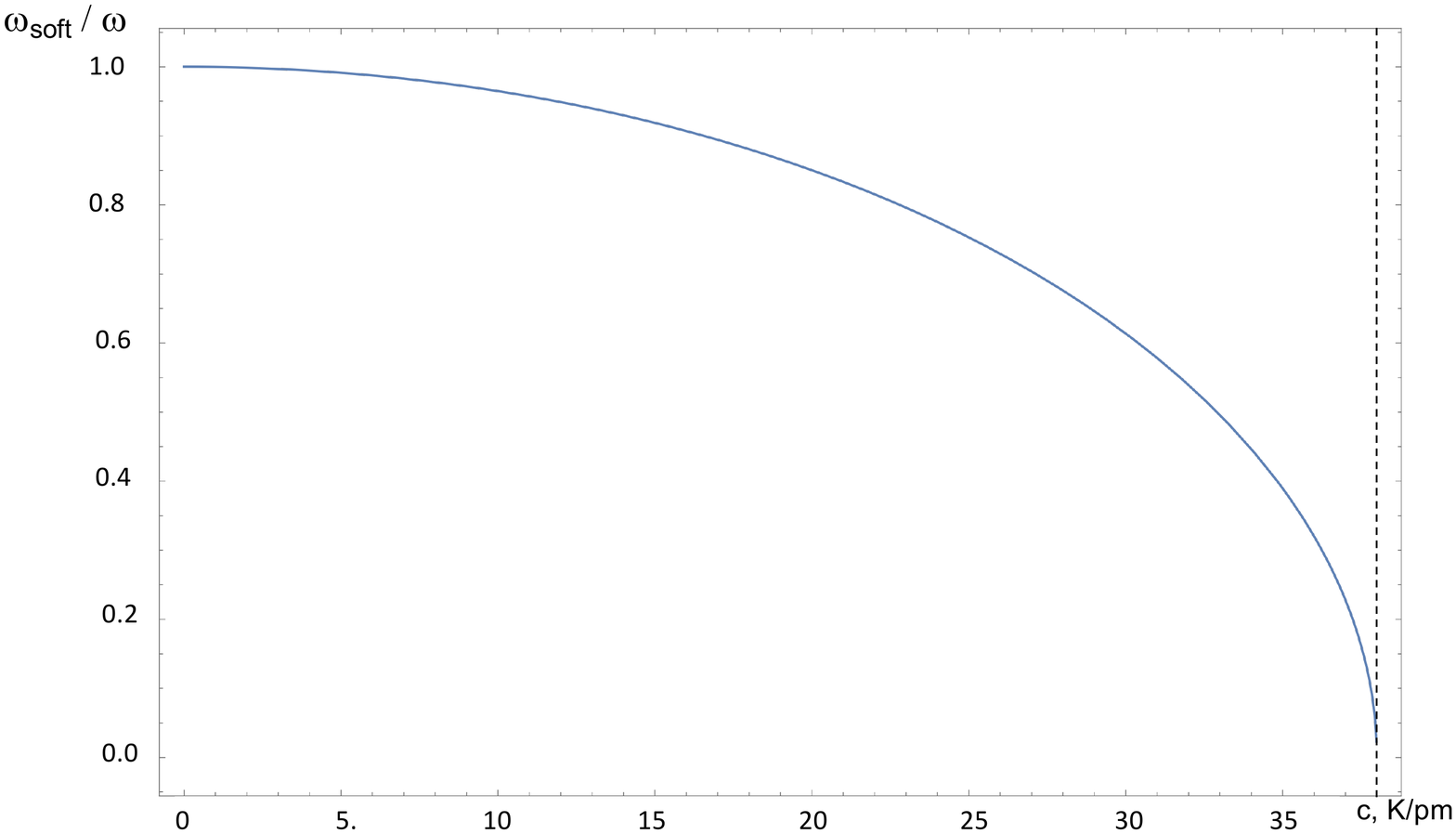}
\caption{\it{Relative softening of the phonon frequency as a function of c}}
\end{figure}

\end{document}